# Modeling Temperature, Frequency, and Strain Effects on the Linear Electro-Optic Coefficients of Ferroelectric Oxides

Yang Liu[1,*], Guodong Ren[2,*], Tengfei Cao[3], Rohan Mishra[3,2,a)], and Jayakanth Ravichandran[1,4,b)]

**AFFILIATIONS**

[1]Mork Family Department of Chemical Engineering and Material Science, University of Southern California, Los Angeles, CA 90089
[2]Institute of Materials Science & Engineering, Washington University in St. Louis, St. Louis, Missouri 63130
[3]Department of Mechanical Engineering & Materials Science, Washington University in St. Louis, St. Louis, Missouri 63130.
[4]Ming Hsieh Department of Electrical Engineering, University of Southern California, Los Angeles, CA 90089

[a)] rmishra@wustl.edu

[b)] j.ravichandran@usc.edu

*These authors contributed equally.

## Abstract

An electro-optic modulator offers the function of modulating the propagation of light in a material with electric field and enables seamless connection between electronics-based computing and photonics-based communication. The search for materials with large electro-optic coefficients and low optical loss is critical to increase the efficiency and minimize the size of electro-optic devices. We present a semi-empirical method to compute the electro-optic coefficients of ferroelectric materials by combining first-principles density-functional theory calculations with Landau-Devonshire phenomenological modeling. We apply the method to study the electro-optic constants, also called Pockels coefficients, of three paradigmatic ferroelectric oxides: $BaTiO_3$, $LiNbO_3$, and $LiTaO_3$. We present their temperature-, frequency- and strain-dependent electro-optic tensors calculated using our method. The predicted electro-optic constants agree with the experimental results, where available, and provide benchmarks for experimental verification.

# Introduction

*AMO*$_3$-type ferroelectric oxides offer strong coupling between electrical, thermal, and optical properties, and enable novel applications that leverage the coupled phenomena. They are currently used in nonvolatile memories, actuators, transducers, and electro-optic (EO) devices, owing to their excellent dielectric, piezoelectric and pyroelectric properties, and optical response. [1–4] For optical applications, ferroelectric oxide perovskites exhibit large EO coefficients with low optical loss, and are the materials of choice for low-power electro-optic devices. Since the 1970s, EO modulators based on LiNbO$_3$ have been used widely in fiber-optic systems due to its good linear EO or Pockels effect ($r_{33}$: 32 pm/V) and high transparency over a large range of wavelength. [5] Thin film deposition of EO oxides, characterization of their optical response, [6] and fabrication of optical devices have undergone significant refinement since the 1990s. [7–10] In line with this development, there is a growing interest in achieving epitaxially grown ferroelectric thin films integrated on silicon-based chips for optical waveguide modulators. [11–16]

Among different ferroelectric oxides, BaTiO$_3$, LiNbO$_3$, and LiTaO$_3$ have been investigated intensively for on-chip EO applications due to their sizable linear EO effect (tetragonal BaTiO$_3$ on Si $r_{42}$: 105 pm/V, $r_{eff}$: 148 pm/V [17] and LiNbO$_3$ on Si $r_{33}$: 17.6 pm/V [18]). However, modeling methods for the EO response of these ferroelectric materials as a function of temperature, frequency, strain and electric dipole orderings has not been well-established. [19–23] In fact, EO effects are shown to be sensitive to the microstructure, and an accurate assessment of this intrinsic property requires single crystals or high-quality thin films, which are not easily accessible or prepared. Therefore, theoretical prediction of the nonlinear optical properties of crystalline materials along with the effect of various experimental conditions, such as strain and temperature, can help to establish performance limits for subsequent experimental verification. In the past decade, sustained efforts on theoretical investigations of nonlinear optical phenomena in oxide perovskites have resulted in accurate methods for predicting these properties. DiDomenico and Wemple revealed the importance of oxygen octahedra in perovskites on their optical properties. [24,25] Ghosez and co-workers calculated the optical susceptibilities, Raman efficiencies, and electro-optic tensors based on density functional perturbation theory. [26–29] More recently, Hamze *et al*., Qiu *et al.,* and Paillard *et al.* studied the effect of strain on the electro-optic tensor. [19–23,30] Furthermore, with the ability to prepare atomically precise heterostructures and superlattices, it is of both scientific and practical importance to understand the mechanism of the EO effect in these complex systems and predict the EO coefficients reliably. [31,32] While first-principles calculation methods used in previous studies are effective in predicting the EO effect for single crystals, modeling EO effects in superlattices and multilayers presents a formidable challenge. The periodicity of the superlattices and multilayers, which span few nm to few 10s of nm, and breadth of phase space in terms of materials and periodicities needed to model EO effects and identify high efficiency structures make first-principles methods computationally expensive and impractical.

Phenomenological models, such as those based on Landau-Devonshire theory [33] enable fast, accurate, and highly scalable calculations of the functional properties of complex structures. It is important to note that Landau-Devonshire model uses input from experimental results or first-principles calculations to fit the coefficients used in the model. Hence, the accuracy of Landau-Devonshire expansion coefficients in subsequent estimation of functional properties is determined by these inputs. For a multicomponent system, such as superlattices and multilayers, one can simulate their physical properties by summing up the thermodynamic free energies of each component as a function of strain, electric fields, and their gradients. [34,35] This approach has been extensively applied for the simulation of dielectric and piezoelectric responses of ferroelectric materials and multilayer heterostructures. [36–38]

The objective of this study is to establish a semi-empirical model to simulate the EO behavior of perovskite ferroelectrics. This model uses the phenomenological Landau-Devonshire model with parameters obtained from first-principles calculations to improve the scalability of EO calculations for complex structures without compromising on speed and accuracy. We show that this model can be applied to prototypical ferroelectric oxides such as $LiNbO_3$, $LiTaO_3$, and $BaTiO_3$. We obtained the free-energy landscape associated with the transition between ferroelectric and paraelectric phases using density-functional theory (DFT) calculations. We extracted Landau-Devonshire coefficients using a polynomial fitting to the energy landscape and calculated the EO coefficients at room temperature for these three prototypical ferroelectric oxides. We find that our model can predict EO coefficients that have good agreement with experimental results, wherever available. Using this model, we have calculated the temperature-dependence of the EO coefficients for LiNbO$_3$ and BaTiO$_3$ and find them to be within 30% of experimental results for most cases. Moreover, the strain effect on the EO coefficient is discussed in the range of -5 to 5% misfit strain for BaTiO$_3$. Our model is able to capture the ferroelectric to paraelectric phase transition, which is associated with a divergence of the EO tensor.

## Methods

**Density-Functional Theory Calculations**

The landscapes of free energy for the different $AM$O$_3$ oxides were computed using DFT as implemented in Vienna Ab-initio Simulation Package (VASP). [39] We used projector augmented-wave (PAW) potentials. [40] In general, the accuracy in the estimation of ferroelectric properties is sensitive to the adopted exchange-correlation functionals such as the local density approximation (LDA), [41] and the semi-local generalized gradient approximation (GGA) in the standard from of Perdew-Burke-Ernzerhof (PBE) [42]. GGA is known to suffer from the so-called super-tetragonality error, which significantly overestimates the structural distortion in conventional perovskite ferroelectrics [43]. For the most-studied oxide ferroelectrics, BaTiO$_3$ and PbTiO$_3$, the lattice distortion, spontaneous polarization, and lattice dynamics predicted by LDA functional agree well with the experimental results. [44] Therefore, we chose LDA to describe the electronic exchange-correlation interactions. We have considered three paradigmatic $AM$O$_3$ oxides, BaTiO$_3$ (*P*4*mm*, *Amm*2, *R*3*m*), LiNbO$_3$ (*R*3*c*), and LiTaO$_3$ (*R*3*c*), for determining structural transition and ferroelectric polarization. A cutoff energy of 700 eV was used to determine the number of planewave basis sets in the calculations. We used Γ-centered 10×10×10 *k*-points mesh for sampling the Brillouin zone of BaTiO$_3$ and 10×10×4 *k*-points mesh for LiNbO$_3$ and LiTaO$_3$. The crystal structures were fully optimized until residual forces were less than $10^{-3}$ eV/Å. The spontaneous polarization induced by the polar soft-phonon modes was calculated based on the modern theory of polarization [45], which is a sum over the contribution from the ionic and electronic charges. Symmetry and distortion-mode analyses were conducted using programs from the Bilbao crystallographic server. [46] The intermediate images corresponding to soft-phonon distortion were interpolated using the ISOTROPY software suite. [47]

**Density-Functional Perturbation Theory Calculations**

We also calculated the EO tensor of the ground-state $R3m$ phase of BaTiO$_3$ purely from first-principles as a comparison to that obtained using the Landau-Devonshire model. The theoretical framework developed by Veithen et al. [26–29] for the computation of EO response under a static or low-frequency electric field perturbation has been implemented in the ABINIT software package. [48,49] Teter extended norm-conserving pseudopotentials [19–23] for BaTiO$_3$ system were used for these calculations and the exchange-correlation interactions were described within LDA. [50] We used 12×12×12 $k$-points mesh and 55 Hartree cutoff energy for all the calculations. To study the effect of strain on EO response, we adopted the same strategy described by Fredrickson *et al*. [21] Varying epitaxial strains between –2 to +2 % with the negative values denoting compressive strain were applied to $a$ and $b$ lattice constants. The optimal $c$ lattice constant for a given epitaxial strain was calculated using the elastic constants of tetragonal BaTiO$_3$ ($C_{11}$ = 222 GPa, $C_{12}$ = 108 GPa, $C_{13}$ = 111 GPa, $C_{33}$ = 151 GPa). [51] The ionic positions in the strained lattices were optimized until the forces were less than 1×10$^{-5}$ eV/A.

**Landau-Devonshire Model**

Landau phenomenological theory is widely used to describe phase transitions and temperature dependence of physical properties of ferroelectrics. [52] Here, we use Helmholtz free energy to describe the thermodynamics due to the convenience in choosing the internal variables: polarization ($P$) and strain ($S$) as independent variables, whereas the electric field ($E$) and the stress are external applied variables. The free energy and free energy density in this article refer to Helmholtz free energy and Helmholtz free energy density, unless noted otherwise. The Helmholtz free energy density ($f_0$) of a ferroelectric system under no external field can be written as an expansion of the order parameter - the polarization ($P$), as: [53]

$$f_0 = a_1(P_1^2 + P_2^2 + P_3^2) + a_{11}(P_1^4 + P_2^4 + P_3^4) + a_{12}(P_1^2 P_2^2 + P_1^2 P_3^2 + P_2^2 P_3^2) +$$
$$a_{111}(P_1^6 + P_2^6 + P_3^6) + a_{112}(P_1^4(P_2^2 + P_3^2) + P_2^4(P_1^2 + P_3^2) + P_3^4(P_1^2 + P_2^2)) +$$
$$a_{123} P_1^2 P_2^2 P_3^2 , \quad (1)$$

where the subscripts *1,2,3* refer to [100], [010], and [001] directions in the crystal, $a_i, a_{ij}, a_{ijk}$ are the phenomenological Landau-Devonshire coefficients, and $P_i$ is the polarization along direction $i$. The temperature dependence of ferroelectricity is governed by the coefficient $a_1$ and it is defined as

$$a_1 = (T - T_0)/2\varepsilon_0 C. \quad (2)$$

The other coefficients are all assumed to be temperature independent. Here $T_0$ and C are the Curie-Weiss temperature and constant above which the system transitions to a paraelectric state, and $\varepsilon_0$ is the dielectric constant of free space, respectively. We set $T_0$ to be 388 K in the entire simulation for BaTiO$_3$ and 1480 K and 950 K for LiNbO$_3$ and LiTaO$_3$, which were observed from experiments. [5,54–56] Classical Landau theory ignores the temperature effect on the higher-order coefficients in the expansion. Nevertheless, it is shown that the higher order terms are actually temperature dependent. [57] We include temperature effects in our calculations, and for simplicity, we only consider the temperature dependent $a_1$ in this work. The effect of temperature-dependent high-order terms, such as $a_{11}$, on EO responses will be the target of future work.

**Model Fitting and Parameters**

The ferroelectric transition from a centrosymmetric reference can be expressed as the result of ionic displacements along a specific direction with charge separation leading to a net electrical dipole moment. [58] By interpolating the ionic displacements from a centrosymmetric structure to a polar phase, the energy as a function of ionic displacements can be mapped using DFT calculations. As has been shown recently by Paoletta and Demkov [19], phonons causing the ionic displacements will in turn alter the electronic energy of the system, and this is the origin of electron-phonon interactions under the adiabatic approximation. That is to say, our DFT calculations for free energy landscape of each distortion mode also reflects the electron-phonon interactions. For the subsequent Landau-Devonshire fittings, we have converted the ionic displacements into spontaneous polarization based on the modern theory of polarization. [59] The landscape of the change in free energy density (J/m$^3$) for the three ferroelectric phase transitions from paraelectric BaTiO$_3$ (*P4/mmm*) as a function of the electric polarization are shown in Figure 1. By fitting the Landau-Devonshire expansion to the change in energy density with polarization, quadratic and higher-order coefficients of the polynomial can be derived for ferroelectric transition along [001], [011], and [111] direction for tetragonal (*P4mm*), orthorhombic (*Amm*2), and rhombohedral (*R3m*) structures, respectively. The free energy density with respect to the polarization $P_{001} = P_3$, $P_{011} = \sqrt{P_2^2 + P_3^2}$, and $P_{111} = \sqrt{P_1^2 + P_2^2 + P_3^2}$ can be described by the following equations, respectively: [60]

$$f_{001} = a_1 P_{001}^2 + a_{11} P_{001}^4 + a_{111} P_{001}^6, \qquad (3.a)$$

$$f_{011} = a_1 P_{011}^2 + a_{11}^O P_{011}^4 + a_{111}^O P_{011}^6, \qquad (3.b)$$

$$f_{111} = a_1 P_{111}^2 + a_{11}^R P_{111}^4 + a_{111}^R P_{111}^6, \tag{3.c}$$

where the superscripts O and R indicate the orthorhombic and rhombohedral phase for BaTiO$_3$, $a_{11}^O = \frac{1}{2}a_{11} + \frac{1}{4}a_{12}$, $a_{111}^O = \frac{1}{4}(a_{111} + a_{112})$, $a_{11}^R = \frac{1}{3}(a_{11} + a_{12})$, and $a_{111}^R = \frac{1}{27}(3a_{111} + 6a_{112} + a_{123})$. We used the "Curve Fitting Toolbox" in MATLAB to fit the free energy density curves obtained from DFT calculations. We fitted the energy density landscape of tetragonal BaTiO$_3$ with the eqn. (3.a) to obtain the $a_1$, $a_{11}$, and $a_{111}$. To get the $a_{12}$ and $a_{112}$, the orthorhombic energy density is fitted to the eqn. (3.b). $a_{123}$ is derived by fitting the energy density of rhombohedral phase using the eqn. (3.c) using all the other parameters obtained from the previous steps. Then all the parameters are manually tuned to minimize the coefficients of determination ($R^2$) of three equations (3, a-c) by slightly changing only one parameter at a time while fixing all remaining parameters. Thus, the whole sets of the Landau-Devonshire coefficients can be derived.

For the case of LiNbO$_3$ and LiTaO$_3$, we applied the same procedure as BaTiO$_3$ but simplified it to [001] direction since we are only interested in the most intense EO tensor component -- $r_{33}$. [61] Therefore, we calculated the free energy density curve for $R3c$ LiNbO$_3$ and LiTaO$_3$ as a function of polarization along [001] direction. The Landau-Devonshire coefficients $a_1$ and $a_{11}$ were obtained by fitting the energy density curve to the eqn. (3.a).

Then, we applied the strain and electrostrictive energy terms to the Landau-Devonshire model to investigate the strain-induced phase changes in BaTiO$_3$. The free energy density $f$ of the thin film as a function of polarization and misfit strain $S_m = (a_s - a_f)/a_s$, [36] where $a_s$ is the substrate lattice parameter and $a_f$ is the lattice constant of the film in its bulk form, is given by: [62]

$$f = a_1^*(P_1^2 + P_2^2) + a_3^* P_3^2 + a_{11}^*(P_1^4 + P_2^4) + a_{33}^* P_3^4 + a_{12}^* P_1^2 P_2^2 + a_{13}^*(P_1^2 P_3^2 + P_2^2 P_3^2) + a_{111}(P_1^6 + P_2^6 + P_3^6) + a_{112}(P_1^4(P_2^2 + P_3^2) + P_2^4(P_1^2 + P_3^2) + P_3^4(P_1^2 + P_2^2)) + a_{123} P_1^2 P_2^2 P_3^2 + \frac{S_m^2}{s_{11}+s_{12}}, \tag{4}$$

where
$$a_1^* = a_1 - \frac{Q_{11}+Q_{12}}{s_{11}+s_{12}} S_m, \tag{4.a}$$

$$a_3^* = a_1 - \frac{2Q_{12}}{s_{11}+s_{12}} S_m, \tag{4.b}$$

$$a_{11}^* = a_{11} + \frac{1}{2}\frac{(Q_{11}^2+Q_{12}^2)s_{11}-2Q_{11}Q_{12}s_{12}}{s_{11}^2-s_{12}^2}, \tag{4.c}$$

$$a_{33}^* = a_{11} - \frac{Q_{12}^2}{s_{11}+s_{12}}, \tag{4.d}$$

$$a_{12}^* = a_{12} - \frac{(Q_{11}^2+Q_{12}^2)s_{12}-2Q_{11}Q_{12}s_{11}}{s_{11}^2-s_{12}^2} + \frac{Q_{44}^2}{2s_{44}}, \tag{4.e}$$

$$a_{13}^* = a_{12} + \frac{Q_{12}(Q_{11}+Q_{12})}{s_{11}+s_{12}}, \tag{4.f}$$

where $Q_{ij}$ are the electrostriction coefficients and $s_{ij}$ are the elastic compliances. The $Q_{ij}$ and $s_{ij}$ values in Table 1 for BaTiO$_3$ are taken from Ref. 63. [63]

Table 1. Elastic compliance ($s_{ij}$) and electrostrictive coefficients ($Q_{ij}$) of BaTiO$_3$ taken from Ref. 63.

| | |
|---|---|
| $s_{11}$ (10$^{-12}$ m$^2$/N) | 8.33 |
| $s_{12}$ (10$^{-12}$ m$^2$/N) | -2.68 |
| $s_{44}$ (10$^{-12}$ m$^2$/N) | 9.24 |
| $Q_{11}$ (m$^4$/c$^2$) | 0.10 |
| $Q_{12}$ (m$^4$/c$^2$) | -0.034 |
| $Q_{44}$ (m$^4$/c$^2$) | 0.029 |

The variation of the free energy density under external electric field is written as:

$$\Delta f = f_0 - E_1 P_1 - E_2 P_2 - E_3 P_3, \tag{5}$$

where $E_1$, $E_2$, and $E_3$ is the applied electric field along $x$, $y$, and $z$ principal crystallographic directions, respectively. The equilibrium configuration is determined by finding the minima of $\Delta f$, where we shall have $\frac{\partial \Delta f}{\partial P} = 0$. Then, the electric field $E$ as a function of polarization can be determined by:

$$E_1 = \frac{\partial f_0}{\partial P_1}, \tag{6.a}$$

$$E_2 = \frac{\partial f_0}{\partial P_2}, \tag{6.b}$$

$$E_3 = \frac{\partial f_0}{\partial P_3}. \tag{6.c}$$

In this article, every time an external electric field is applied, we solve eqns. (6.a-c) to deduce the field-induced polarizations. Then the obtained polarizations are applied to solve the corresponding eqns. (8-9) in the following paragraphs. In this way, it will always maintain the thermodynamic equilibrium: $\frac{\partial \Delta f}{\partial P} = 0$.

The dielectric tensor $\varepsilon_{ij}$ can be defined in terms of the first-order derivative of polarization with respect to the external electric field. Here, we summarize the derived dielectric constants for tetragonal ($P_1 = P_2 = 0, P_3 \neq 0$) and orthorhombic (orthorhombic, $P_1 = P_2 \neq 0, P_3 = 0$) BaTiO$_3$ phases in the box I and II. We don't include the low-temperature rhombohedral phase since the rhombohedral phase is not accessible in the experiments though strain engineering at room temperature.

$$\varepsilon_{11} = \varepsilon_{22} = \frac{1}{2a_1^* + 2a_{13}^* P_3^2 + 2a_{112} P_3^4},$$

$$\varepsilon_{33} = \frac{1}{2a_3^* + 12a_{33}^* P_3^2 + 30a_{111} P_3^4}.$$

Box I. Expression of spontaneous polarization and dielectric constants for tetragonal BaTiO$_3$.

$$X_{11} = 2a_1^* + 12a_{12}^* P_1^2 + 2a_{12}^* P_2^2 + 30a_{111} P_1^4 + a_{112}(12P_1^2 P_2^2 + 2P_2^4),$$

$$X_{22} = 2a_1^* + 12a_{12}^* P_2^2 + 2a_{12}^* P_1^2 + 30a_{111} P_2^4 + a_{112}(12P_1^2 P_2^2 + 2P_1^4),$$

$$X_{12} = 4a_{12}^* P_1 P_2 + 8a_{112}(P_1^3 P_2 + P_1 P_2^3),$$

$$X_{33} = 2a_3^* + 2a_{13}^*(P_1^2 + P_2^2) + 8a_{112}(P_1^3 P_2 + P_1 P_2^3),$$

$$\varepsilon_{11} = \frac{X_{22}}{X_{11}X_{22} - X_{12}^2}, \quad \varepsilon_{22} = \frac{X_{11}}{X_{11}X_{22} - X_{12}^2}, \quad \varepsilon_{33} = \frac{1}{X_{33}}.$$

Box II. Expression of spontaneous polarization and dielectric constants for orthorhombic BaTiO$_3$.

The propagation of light in a crystal is determined by the refractive index $n_{ij}$. The relation between the dielectric constant and the refractive index is $n_{ij}^2 = \varepsilon_{ij}/\varepsilon_0$. The linear EO tensor $r_{ijk}$ describes the change of refractive index of a crystal in response to the applied electric field. Therefore, we write the linear EO tensor $r_{ijk}$ as first-order dependence of the inverse of refractive index square when a static or low-frequency modulating electric field $E_k$ is applied:

$$\Delta(n_{ij}^{-2}) = r_{ijk} E_k \quad (7)$$

The index *ijk* refers to the *ij* component of the refractive index and the dielectric tensor, for an applied electric field along the *k* direction. [20,28,64] For the following paragraphs, we denote the index *ij* with Voigt notations, i.e. $11 \to 1, 22 \to 2, 33 \to 3, 23 \to 4, 13 \to 5,$ and $12 \to 6$.

From eqn. (6), we have the electric field as a function of polarization, and equations in the box I and II give the dielectric constant as a function of polarization by substituting them into eqn. (7). Thus, given all the Landau-Devonshire coefficients obtained using polynomial fitting, the EO coefficients can be obtained. Here, we consider the case of tetragonal and orthorhombic phases of BaTiO$_3$, as examples. The EO tensors in the ferroelectric tetragonal *P4mm* phase of BaTiO$_3$ have three independent elements (Voigt notations), $r_{13}$, $r_{33}$, and $r_{42}$. [64]

$$r_{13} = \frac{\varepsilon_0(4a_{12}^*P_3+8a_{112}P_3^3)}{2a_1^*+12a_{11}^*P_3^2+30a_{111}P_3^4}, \tag{8.a}$$

$$r_{33} = \frac{\varepsilon_0(24a_{11}^*P_3+120a_{111}P_3^3)}{2a_1^*+12a_{11}^*P_3^2+30a_{111}P_3^4}, \tag{8.b}$$

$$r_{42} = \varepsilon_0\left(\frac{8a_{123}P_3}{4a_{12}^*+4a_{123}P_3^2}+\frac{4a_{13}^*P_3+8a_{112}P_3^3}{2a_1^*+2a_{13}^*P_3^2+2a_{112}P_3^4}+\frac{4a_{13}^*+24a_{112}P_3^2}{4a_{13}^*P_3+8a_{112}P_3^3}\right). \tag{8.c}$$

For the derivation details, please see the appendix.

The orthorhombic phase of BaTiO$_3$ is not a thermodynamically stable phase at room temperature. However, it could be stabilized under tensile strain, such as epitaxially grown orthorhombic BaTiO$_3$ films on MgO. [7] The EO tensors of orthorhombic BaTiO$_3$ are

$$r_{13} = \frac{\varepsilon_0(24a_{11}^*P_1+120a_{111}P_1^3+24a_{112}P_1P_2^2+4a_{12}^*P_2+24a_{112}P_2P_1^2+8a_{112}P_2^3)}{(2a_3^*+2a_{13}^*(P_1^2+P_2^2)+2a_{112}(P_1^4+P_2^4)+2a_{123}(P_1^2P_2^2))}, \tag{9.a}$$

$$r_{33} = \frac{\varepsilon_0(4a_{13}^*(P_1+P_2)+8a_{112}(P_1^3+P_2^3))}{(2a_3^*+2a_{13}^*(P_1^2+P_2^2)+2a_{112}(P_1^4+P_2^4)+2a_{123}P_1^2P_2^2)}, \tag{9.b}$$

$$r_{42} = \varepsilon_0\left(\frac{4a_{13}^*P_1+8a_{112}P_1^3}{2a_1^*+12a_{11}^*P_1^2+2a_{12}^*P_2^2+30a_{111}P_1^4+12a_{112}P_1^2P_2^2+2a_{112}P_2^4}+\frac{4a_{13}^*P_2+8a_{112}P_2^3}{4a_{12}^*P_1P_2+8a_{112}P_1^3P_2+8a_{112}P_1P_2^3}\right). \tag{9.c}$$

It is worth noting that the $a_1$ coefficient is temperature-dependent, as shown in eqn. (2). Hence, the temperature-dependent EO responses could also be obtained using this method.

For LiNbO$_3$ and LiTaO$_3$, the EO coefficient $r_{33}$ is

$$r_{33} = \frac{\varepsilon_0 24a_{11}P_3}{2a_1+12a_{11}P_3^2}. \tag{10}$$

Microscopically, there are three contributions induced by a modulating electric field to the EO tensor: the electronic contribution from the valence electrons, the ionic contribution from the displacement of the ions, and the piezoelectric contribution from the distortion of the unit cell through the converse piezoelectric effect. [19-25] At moderately high modulating frequencies, that are low compared to the optical phonon modes, ionic contributions to the EO tensor dominate. At such frequencies, strain relaxations can be avoided. To investigate the frequency dispersion of the coefficients, we also applied the time-dependent Ginzburg-Landau (TDGL) equation [6,33]:

$$\frac{\partial P_i(t)}{\partial t} = -L \frac{\partial F(P_i)}{\partial P_i} \quad (11)$$

where $L$ is the kinetic coefficient (proportional to the dipole motion velocity) and $t$ is time. The energy function $F$ is $\Delta f$ in equation (5) except the fact that the applied electric field is static but here, dynamic electric field is used as a triangle wave function:

$$E(t) = E_0 \sin^{-1}[\sin(f\pi t)] \quad (12)$$

where $E_0$ is the amplitude of the electric field and $f$ the frequency.

It is not easy to solve TDGL explicitly, as we have done for the static calculations. Therefore, we performed the calculations using finite element method to obtain the EO coefficients from 10 Hz to 1 GHz. In this work, we provide an example of the frequency dependent $r_{33}$ for the tetragonal BaTiO$_3$. We have ignored optical frequencies where the electronic contributions dominate, as they are not included in our model. It is acceptable to ignore the electronic contributions at low frequencies, as they have been shown to be relatively small compared to the dipole contributions in perovskites [20,28].

In summary, we have extracted the Landau-Devonshire coefficients from the free-energy landscape calculated using first-principles DFT using equations (2.a-c). We have simulated the dependence of polarization on the applied electric field using equations (6.a-c) and calculated the dielectric constant using the relations shown in Box I and II. The electric field and dielectric constant as a function of polarization are then plugged into equation (7) to obtain the electro-optic tensors. In this case, we give two solutions for tetragonal and orthorhombic BaTiO$_3$ in equations (8.a-c) and (9.a-c), respectively.

## Results and Discussion

We have used the LDA functional to calculate the Helmholtz free energy density as a function of the polarization for LiNbO$_3$, LiTaO$_3$, and BaTiO$_3$. In the case of BaTiO$_3$, the high-temperature phase has a centrosymmetric cubic structure. However, as the temperature decreases, a sequence of phase transitions are observed experimentally as follows: cubic $\xrightarrow{388K}$ tetragonal $\xrightarrow{273K}$ orthorhombic $\xrightarrow{183K}$ rhombohedral. [54] These three ferroelectric phase transitions result in a change in the direction of the spontaneous polarization from the [001] axis (tetragonal, $P_1 = P_2 = 0, P_3 \neq 0$), to the [110] axes (orthorhombic, $P_1 = P_2 \neq 0, P_3 = 0$), and to the [111] axes (rhombohedral, $P_1 = P_2 = P_3 \neq 0$) as the temperature decreases. In Figure 1, the energy density changes for the phase transition from the cubic structure directly to tetragonal, orthorhombic, and rhombohedral phase of BaTiO$_3$ are shown. The spontaneous polarization is corresponding to the polarization value, where the energy reaches the minima at the bottom of the double-well. The calculated spontaneous polarization values for tetragonal, orthorhombic, and rhombohedral BaTiO$_3$ are 0.24 *C/m²*, 0.27 *C/m²*, and 0.32 *C/m²*, respectively.

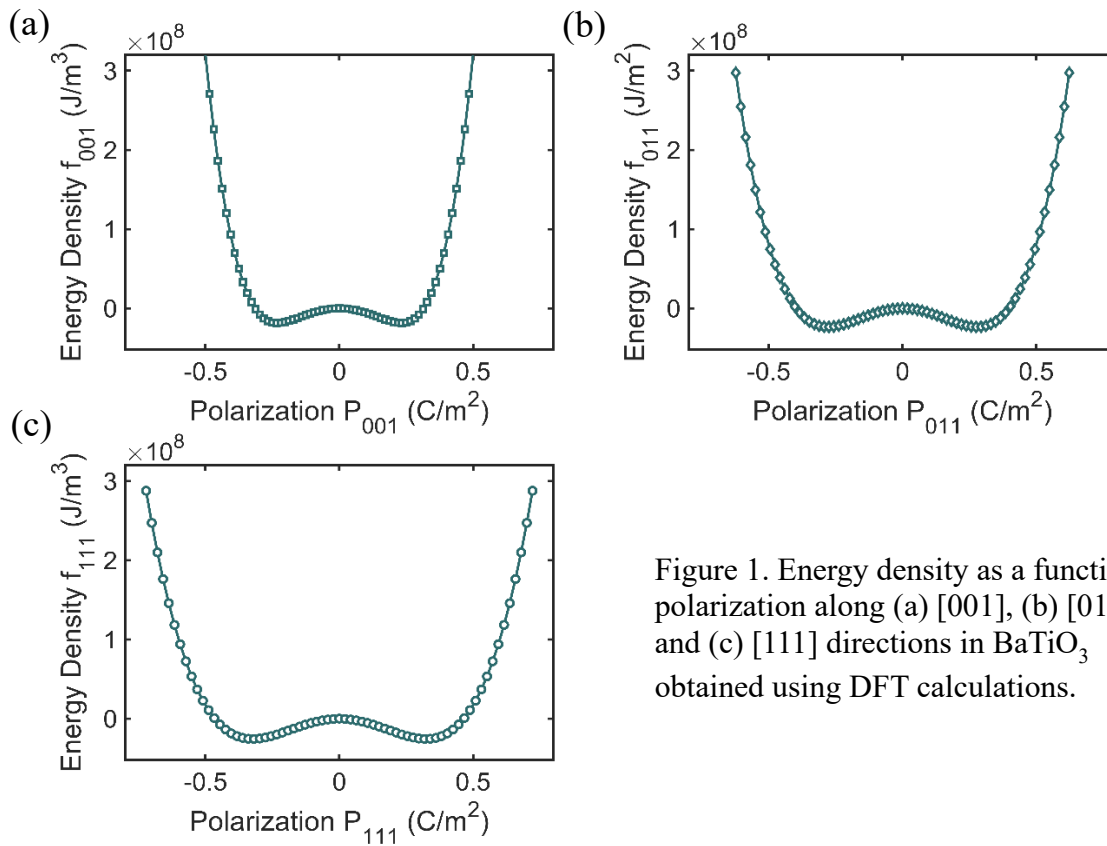

Figure 1. Energy density as a function of polarization along (a) [001], (b) [011], and (c) [111] directions in BaTiO$_3$ obtained using DFT calculations.

In the case of LiNbO$_3$ and LiTaO$_3$, both undergo a transition from a high temperature rhombohedral paraelectric $R\bar{3}c$ phase to a low temperature ferroelectric $R3c$ phase at 1480 $K$ and 950 $K$, respectively. [5,55,56] At room-temperature, the spontaneous polarization points along the $c$-axis direction ($P_1 = P_2 = 0, P_3 \neq 0$) for the ferroelectric rhombohedral phase of LiNbO$_3$ and LiTaO$_3$. We have compared the change in Helmholtz free energy density with respect to the polarization along [001] direction for BaTiO$_3$, LiNbO$_3$ and LiTaO$_3$, as shown in Figure 2. The double-well depth here is a quantitative indicator of the energetic stability of the ferroelectric phase with respect to the paraelectric phase. Tetragonal BaTiO$_3$ yields a shallow double-well indicating a relatively easier transition from the ferroelectric to paraelectric phase.

Table 2. Extracted Landau-Devonshire coefficients from the DFT calculated free energy curves.

| Landau-Devonshire Coefficient | BaTiO$_3$ | | LiNbO$_3$ | | LiTaO$_3$ | |
| --- | --- | --- | --- | --- | --- | --- |
| | This work | Ref. 67 | This work | Ref. 67 | This work | Ref. 67 |
| $a_1$ (Nm$^2$/C$^2$) | -6.07×10$^8$ | -4.74×10$^7$ | -1.20×10$^9$ | -6.28×10$^8$ | -1.54×10$^9$ | -1.006×10$^9$ |
| $a_{11}$ (Nm$^6$/C$^4$) | 4.32×10$^9$ | -2.10×10$^8$ | 9.03×10$^8$ | 1.26×10$^9$ | 2.21×10$^9$ | 9.01×10$^8$ |
| $a_{12}$ (Nm$^6$/C$^4$) | 6.29×10$^9$ | 7.97×10$^8$ | | | | |
| $a_{111}$ (Nm$^{10}$/C$^6$) | 1.29×10$^{10}$ | 1.29×10$^9$ | | | | |
| $a_{112}$ (Nm$^{10}$/C$^6$) | -1.44×10$^{10}$ | -1.95×10$^9$ | | | | |
| $a_{123}$ (Nm$^{10}$/C$^6$) | -1.67×10$^{10}$ | -2.50×10$^9$ | | | | |

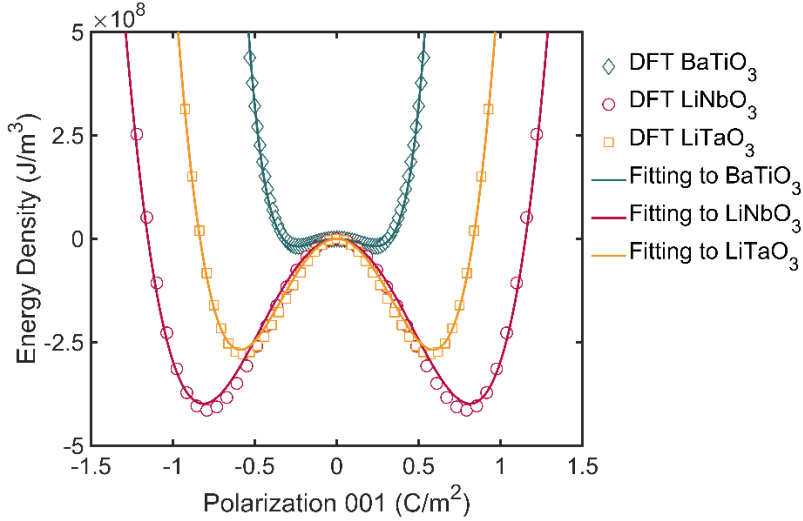

Figure 2. Dependence of free energy density on the polarization along the [001] crystallographic direction in $LiNbO_3$, $LiTaO_3$, and $BaTiO_3$. The solid lines indicated polynomial fitting to the polarization dependent energy curves for $LiNbO_3$, $LiTaO_3$, and $BaTiO_3$ along the [001] axis.

The Landau-Devonshire coefficients are extracted by fitting the double-well energy curves obtained from the first-principles calculations to a polynomial expression (eqn. (1)). Under rigid symmetry framework, the 6$^{th}$ order series expansions are commonly accepted as the basic free energy format describing the ferroelectric phase transitions in $BaTiO_3$. [60] A polynomial with higher-order expansion terms will yield better accuracy but the large number of fitting parameters can also lead to overfitting. Moreover, the higher-order terms require additional information at high polarization region and their physical meaning still remains unclear. [65] Hence, we performed the polynomial fitting to the free energy density curves of $BaTiO_3$ up to 6$^{th}$ order expansions to ensure that the critical aspects of the electro-optic phenomena can be sufficiently described without overfitting. We fit the polynomial of eqns. (3.a), (3.b), and (3.c) to free energy density curves of the three ferroelectric phases: tetragonal, orthorhombic, and rhombohedral, respectively, by simultaneously and manually adjusting the fitting parameters to find the smallest $R^2$ (coefficient of determination). The $R^2$ values are 0.998, 0.999, and 0.996 for tetragonal, orthorhombic, and rhombohedral phase of $BaTiO_3$, respectively. Order parameter just below the Curie temperature has rich information for all the polynomial coefficients, including double P-E loops, [66] while the free energy curve for a system much below the phase transition temperature is a simple double well without much information about higher order polynomials to the expansion coefficients, which is the case for LiNbO$_3$ and LiTaO$_3$. [5,56] The fourth-order polynomial can already provide a good fit to the first-principles data, where $R^2$ is already 0.999. Hence, the higher-order terms were omitted for LiNbO$_3$ and LiTaO$_3$. The Landau-Devonshire coefficients extracted for the noted materials are presented in Table 2. The coefficients from Long-Qing Chen [67] are listed in the table as well for comparison.

We now apply the previously fitted parameters to calculate the electro-optic coefficients using equations (8.a-c). Figure 3 shows the comparison between the calculated values using the model and experimental results. [61] Overall, our method predicts the sign of the EO constants correctly and the values of the EO coefficients are in good agreement with the experimental values. The calculated $r_{33}$ for LiNbO$_3$ is in close agreement (deviation is ~ 1.8%) with the experimentally reported results. [61] The calculated values of $r_{33}$ are ~30% larger than the experimental value of tetragonal BaTiO$_3$ and LiTaO$_3$. Our theoretical $r_{13}$ value of BaTiO$_3$ also matches well with the experimental ones. [61] However, the calculated $r_{42}$ is 88% lower than the experimentally reported value. [61] The deviation from experimental values could be due to extrinsic factors such as stoichiometry and structural quality of the samples and domain structures. Furthermore, DFT calculations using the LDA functional underestimate the polar distortion from paraelectric phases, which results in relatively shallow double-well depths. [44] Therefore, the accuracy of Landau-Devonshire fitting parameters could be further improved by applying more reliable exchange-correlation functionals, such as the recently developed strongly constrained and appropriately normed (SCAN) meta-GGA functional, which has been shown to systematically improve over LDA for structural properties and ferroelectric transitions of diversely bonded materials. [68] These advances will be part of future work.

The temperature dependence of the EO coefficients is another way to examine the accuracy of the model. We compare the calculated results with the experimental values for the LiNbO$_3$ in Figure 4 (a). [69] The model captures the experimentally observed increase in $r_{33}$; however, it consistently over-estimates the value with a larger deviation on decreasing the temperature. As temperature approaches 0 K, the deviation is < 20%, and, as the temperature reaches room temperature, the deviation narrows down to < 10%. To the best of our knowledge, the experimental results of temperature dependency of $r_{33}$ for $LiTaO_3$ and $BaTiO_3$ have not been reported in the literature, and hence, we could not make this comparison for these two compounds. Nevertheless, we compare our results (noted as green hollow circles with a trendline) with the two available first-principles results from Veithen *et al.* [27] and Pietro *et al.* [70] for tetragonal $BaTiO_3$, as depicted with red and yellow markers in Figure 3 (b), respectively. The electro-optic coefficient $r_{33}$ increases with temperature below the Curie temperature and shows a divergent trend close to the Curie temperature, as the spontaneous polarization abruptly drops to zero at and above the Curie temperature.

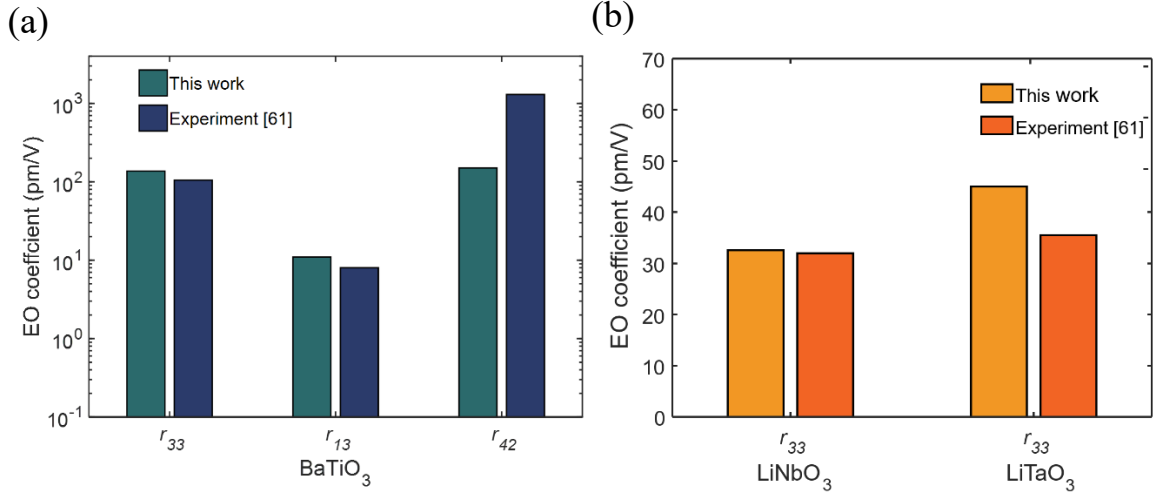

Figure 3. Theoretical and experimental (Ref. [61]) electro-optic coefficients of (a) tetragonal $BaTiO_3$, (b) $LiNbO_3$ and $LiTaO_3$.

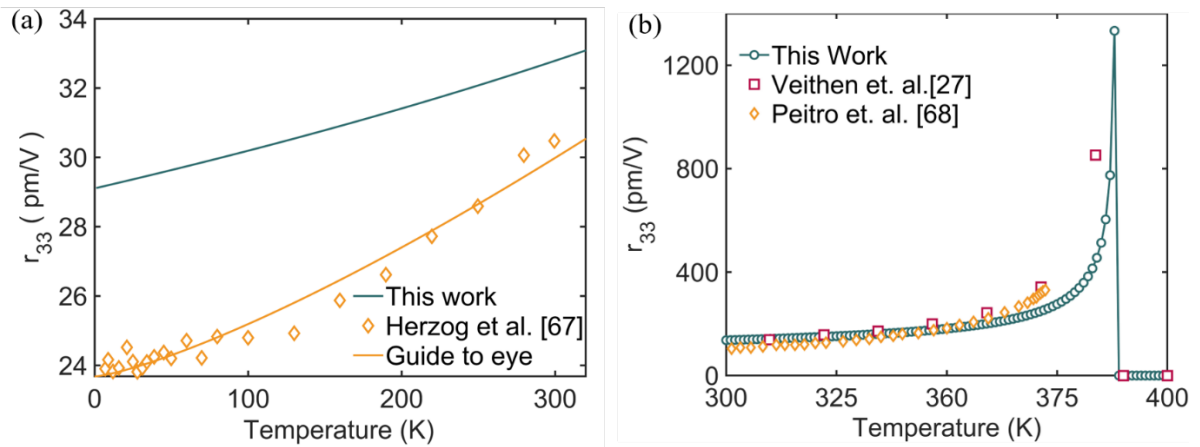

Figure 4. Temperature dependence of electro-optic coefficients in (a) $LiNbO_3$ and (b) $BaTiO_3$.

It is worth reemphasizing that in our model, we assume $a_1$ as the only temperature-dependent term and hence, the temperature dependence of EO response is essentially attributed to it. The contribution of other terms to EO coefficients in equation (8.a) and (9.a), such as $a_{11}P_3^2$, are typically much smaller than $a_1$. Hence, we resorted to this simplification. Any corrections to this simplification will be explored in the future depending on the availability of the experimental results. EO coefficients tend to be extremely large as the temperature approaches the Curie temperature, $T_0$. By definition, $a_1$, given in eqn. (2), converges to 0 as the temperature reaches $T_0$. This hints that small $a_1$ is desirable to achieve a large EO coefficient. *To obtain a small $a_1$, the energy barrier for switching the polarization from one energy well to the other has to be low. It indicates that the origin of this electro-optic enhancement is attributed to the ease of the ferroelectric switching as manifested in the free energy landscape.* From Figure 2 and Table 1, BaTiO$_3$ has the shallowest energy well and the smallest absolute value of $a_1$, which leads to the largest $r_{33}$ among the three ferroelectric oxides. This also explains why the EO coefficients of relaxor ferroelectric oxide alloys are high, for which the corresponding ferroelectric switching energy barrier is relatively low. [31]

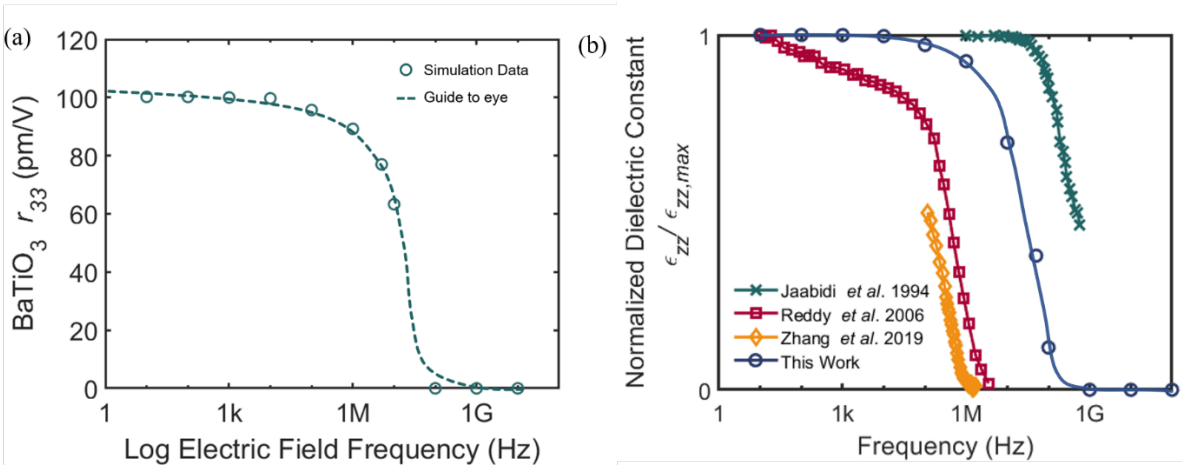

Figure 5. (a) Dependence of the electro-optic coefficient, $r_{33}$, in tetragonal BaTiO$_3$ on the frequency of the applied electric field. (b) The frequency-dependent normalized dielectric constant for tetragonal BaTiO$_3$, in comparison with experimental data. [80-82]

The frequency dispersion of $r_{33}$ and the normalized dielectric constant are shown in Figure 5. The EO coefficient $r_{33}$ turns out to be 100 pm/V at 1 kHz, which is the most common frequency for dielectric measurements. Compared to its experimental value of 105 pm/V, the deviation has dropped from 30 % for static-field calculations ($r_{33}$: 137 pm/V) to -5 % for dynamic calculations. The EO coefficient decreases with increasing frequency of the electric field. As the frequency approaches 10 MHz, the EO coefficient is 63.3 pm/V, as shown in Figure 5(a). We believe the magnitude of EO coefficients are limited by the polarization switching velocity and hence, show a decreasing trend with increasing frequency. Typically, the polarization switching includes three stages: nucleation of domains, domain growth, and sideways motion of newly formed domains. [71–74] Besides nucleation and domain motion, polarization switching can also occur with the simultaneous reversal of all local dipoles, as proposed by Landau and Lifshitz. [75] This type of polarization switching is called intrinsic polarization switching or "Landau switching". In this work, we will follow the Landau approach. The switching time for Landau switching used in this work is sub nanoseconds, which is comparable to the fastest domain wall driven switching record (220 ps). [76,77] In comparison, the measured polarization switching time usually varies from µs to ns scale. [78,79] The intrinsic switching is believed to be much faster than domain nucleation and motion. To evaluate the switching time assumption, we also calculate the frequency-dependent dielectric constant across the 10 kHz to 1 GHz range and qualitatively compare it with the experimental results, as shown in Figure 5 (b). The dielectric dispersion from our method falls in between the data from a bulk single domain $BaTiO_3$ single crystal [80] and single crystal thin films data, [81,82] which proves that the selected switching time in this work is reasonable.

In this work, we demonstrated the dipole contribution to the EO coefficient, specifically $r_{33}$ for tetragonal $BaTiO_3$, by varying the frequency of the electric field. Admittedly, we have not considered the electronic contribution in our model, which dominates at even higher frequencies ~ 100 GHz level. This contribution is relatively small compared to the dipole contribution in perovskites, and is acceptable to ignore it. [20,28] In the following, we demonstrate the role of the piezoelectric term to the EO coefficient by applying strain and adding the electrostrictive terms to the free energy.

Figure 6 shows the EO tensor as a function of strain with misfit strains ranging from -5% to 5% along the in-plane *a* and *b* axes for BaTiO$_3$, which was obtained using eqns. 8(a-c) and 9(a-c). The thermodynamically stable phases are obtained by minimizing the total free energy $F$ under a given misfit strain. The calculation was performed by setting the temperature in the free energy expansion coefficient, $a_1$, as room temperature. We obtained two stable single-phase states. The stable strain condition for tetragonal phase is denoted as T with yellow shade ($P_1 = P_2 = 0, P_3 \neq 0$) below -1% compressive strain, and for orthorhombic phase, it is denoted as O with blue shade ($P_1 = P_2 \neq 0, P_3 = 0$) above ~1.9% tensile strain. For the phase region between the T and O phase, it contains a two-phase mixture (T+O). At zero strain state, we obtain almost identical values of electro-optical tensors as bulk values summarized in Figure 4. The small deviation comes from that the electrostrictive energy term is not included in the previous calculations. The strain-induced polarization variations under compression and tensile strain generate a large contribution to the relevant EO constants. We predict a surprisingly high value, up to thousands of pm/V, for $r_{42}$ coefficient of BaTiO$_3$ under ~ $-1.3$% compressive strain and $r_{33} \approx 800$ pm/V under 1.8% tensile strain, which is $1 - 2$ orders of magnitude larger than $r_{33} = 32$ pm/V for LiNbO$_3$. [61] In this specific case, at the transition region between T and T-O, the sign change of $a_1^*$ leads to a large value of the $r_{42}$ coefficient; similarly, $r_{33}$ is large at the T-O and O phase boundary due to the vanishingly small $a_3^*$ values. Similar EO coefficient enhancement has been demonstrated as a function of temperature near a ferroelectric phase transition due to the divergence of the dielectric constant. [24,70] The polarization rotation at the phase boundary is easier to achieve than in a single phase region, as the corresponding switching barrier is small. This could also explain why ferroelectrics at the phase boundary usually have large dielectric, piezoelectric, and electro-optic responses. [83–85]

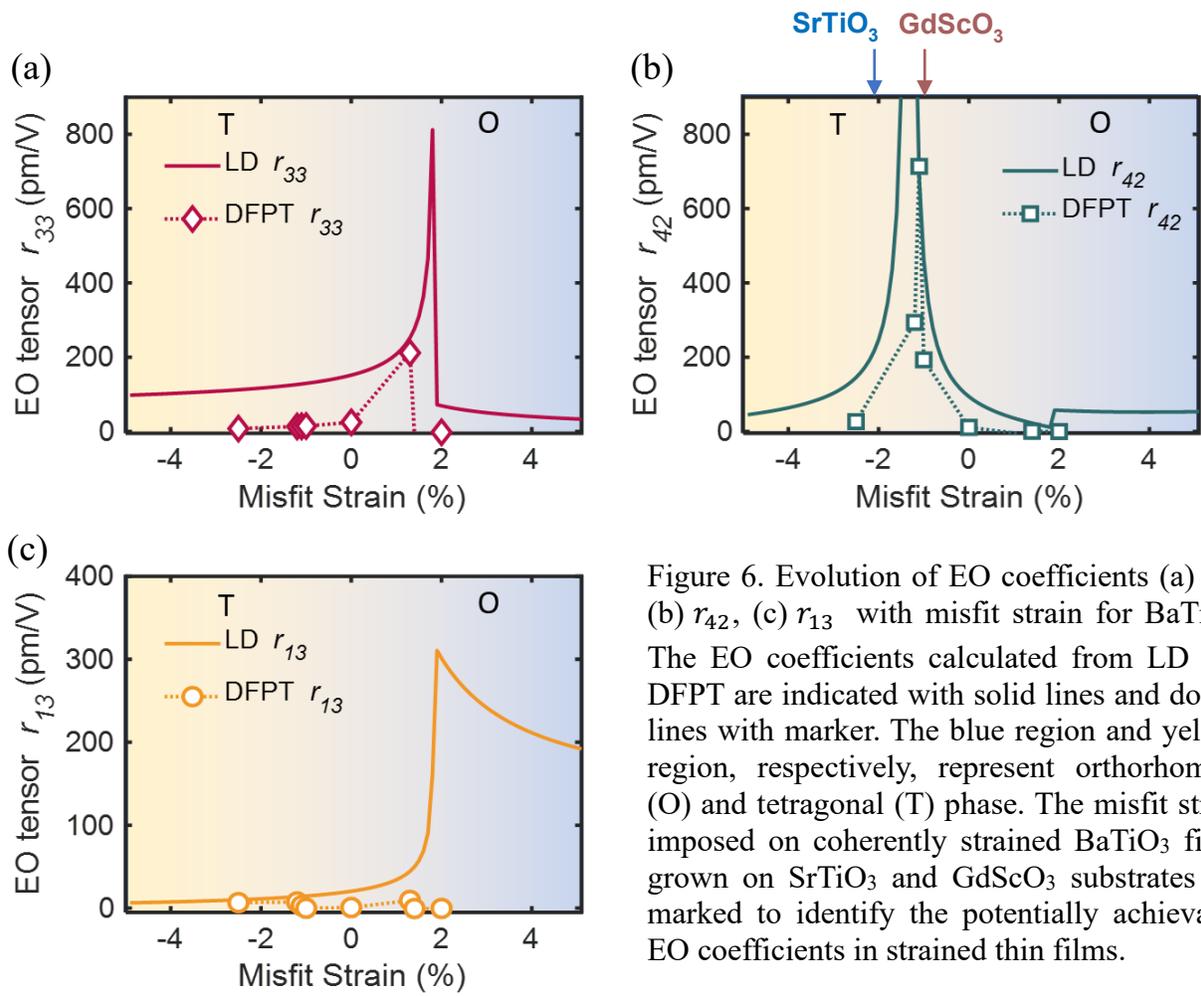

Figure 6. Evolution of EO coefficients (a) $r_{33}$, (b) $r_{42}$, (c) $r_{13}$ with misfit strain for BaTiO$_3$. The EO coefficients calculated from LD and DFPT are indicated with solid lines and dotted lines with marker. The blue region and yellow region, respectively, represent orthorhombic (O) and tetragonal (T) phase. The misfit strain imposed on coherently strained BaTiO$_3$ films grown on SrTiO$_3$ and GdScO$_3$ substrates are marked to identify the potentially achievable EO coefficients in strained thin films.

We have compared the change in EO response of BaTiO$_3$ under strain using our model, with those obtained using computationally intensive first-principles calculations as discussed in the Methods section. We obtain qualitatively similar results, as depicted in Figure 5. The first-principles DFPT results also indicate the divergence of $r_{33}$ and $r_{42}$ at ~1.3% and −1.0% strain, respectively. This clearly demonstrates that our model could qualitatively describe the physics of EO response under epitaxial strain conditions, although the absolute values should be taken with a grain of salt. We give the potential substrates in Figure 5 that one could use to achieve misfit compressive strain close to 1% to obtain large EO coefficients in coherently strained thin films of BaTiO$_3$ using the established strain engineering approach. The realization of low loss thin films in these limits remains a challenge, where intrinsic and extrinsic defects may play an important role. Other factors,

Page 20 of 31

such as the formation and distribution of ferroelectric domains at low fields, any inhomogeneity in application of electric fields will also influence dielectric losses. Overcoming these issues are critical to enable high EO coefficient materials and devices for future photonic technologies.

## Conclusion

A methodology to predict the EO coefficients in ferroelectric oxides as a function of strain, modulation frequency, and temperature is demonstrated here. This method enables highly scalable calculation of EO coefficients by combining computationally expensive, but accurate, first-principles calculations with scalable phenomenological Landau-Devonshire theory. We applied our approach to three representative ferroelectric oxides, namely, LiNbO$_3$, LiTaO$_3$, and BaTiO$_3$. The calculated EO coefficients are in good agreement with the experimental results. And the relevance of specific model parameters for EO effect are discussed. In the light of the previous discussion of temperature and strain effects on the EO responses, we conclude that small $a_1$ ($a_1^*$ for strained case) and shallow energy barrier are favorable for high-$r$ EO materials. One way to reduce $a_1$ is fabrication of ferroelectric/dielectric (FE/DE) heterostructures. As we introduced earlier, Helmholtz free energy is simply the sum of all the energy components in Landau-Devonshire theory. The total $a_1$ is the sum of $a_1$ of the ferroelectric and dielectric phases weighted according to their respective volume fractions. $a_1$ is negative for ferroelectrics and positive for dielectrics. By engineering the volume fraction of each component, one can achieve small $a_1$ coefficients with high EO response. Another aspect of EO performance enhancement in ferroelectrics or multilayer heterostructures is compositional engineering of the phase boundary involving the ferroelectric-paraelectric phases or two different ferroelectric phases such as the O- and T-phases in a mixed-phase BaTiO$_3$ film. Low switching energy at phase boundary makes the polarization switching (between up and down) and/or polarization rotation (between two ferroelectric states) easier and hence, can generate the enhanced EO responses. [86] Our method can be extended to simulate the EO response of FE/DE heterostructures or film having phase boundary, once the complete set of the bulk, elastic, electrostrictive and gradient energy terms are available either from first-principles calculations or from experiments. We expect that this method will pave a way to discover new materials with high electro-optic performance.

# Appendix

The EO tensors, $r_{13}, r_{33}$, and $r_{42}$, are defined in equation 7. Here we provide detail derivations to them. From equation 6.b-c, we have $E_2$ and $E_3$ as

$$E_2 = \frac{\partial f}{\partial P_2} = 2a_1^* P_2 + 4a_{11}^* P_2^3 + 2a_{12}^* P_1^2 P_2 + 2a_{13}^* P_2 P_3^2 + 6a_{111} P_2^5 \tag{12.a}$$

$$+ a_{112}\left(2P_2(P_3^4 + P_1^4) + 4P_2^3(P_1^2 + P_3^2)\right) + 2a_{123} P_1^2 P_2 P_3^2$$

$$E_3 = \frac{\partial f}{\partial P_3} = 2a_3^* P_3 + 4a_{33}^* P_3^3 + 2a_{13}^* P_3(P_1^2 + P_2^2) + 6a_{111} P_3^5 \tag{12.b}$$

$$+ a_{112}\left(2P_3(P_1^4 + P_2^4) + 4P_3^3(P_1^2 + P_2^2)\right) + 2a_{123} P_1^2 P_2^2 P_3$$

For the left hand side of the equation (7), we have:

$$\frac{1}{n_1^2} = \varepsilon_0 \frac{\partial^2 f}{\partial P_1^2} = \varepsilon_0(2a_1^* + 12a_{11}^* P_1^2 + 2a_{12}^* P_2^2 + 2a_{13}^* P_3^2 + 30a_{111} P_1^4 \tag{13.a}$$

$$+ a_{112}(12P_1^2(P_2^2 + P_3^2) + 2(P_2^4 + P_3^4)) + 2a_{123} P_2^2 P_3^2)$$

$$\frac{1}{n_3^2} = \varepsilon_0 \frac{\partial^2 f}{\partial P_3^2} = \varepsilon_0(2a_3^* + 12a_{33}^* P_3^2 + 2a_{13}^*(P_1^2 + P_2^2) + 30a_{111} P_3^4 + a_{112}(12P_3^2(P_1^2 + P_2^2) \tag{13.b}$$

$$+ 2(P_1^4 + P_2^4)) + 2a_{123} P_1^2 P_2^2)$$

$$\frac{1}{n_4^2} = \varepsilon_0 \frac{\partial^2 f}{\partial P_2 \partial P_3} = \varepsilon_0(4a_{13}^* P_2 P_3 + a_{112}(8P_2^3 P_3 + 8P_2 P_3^3) + 4a_{123} P_1^2 P_2 P_3) \tag{13.c}$$

Then substituting the eqns. 12 and eqns. 13 to eqn.7 we obtain:

$$r_{11} = \frac{\partial\left(\frac{1}{n_1^2}\right)}{\partial E_3} = \frac{\varepsilon_0(4a_{12}^* P_3 + 8a_{112} P_3^3)}{2a_1^* + 12a_{11}^* P_3^2 + 30a_{111} P_3^4} \tag{14.a}$$

$$r_{33} = \frac{\partial\left(\frac{1}{n_1^2}\right)}{\partial E_3} = \frac{\varepsilon_0(24a_{11}^* P_3 + 120a_{111} P_3^3)}{2a_1^* + 12a_{11}^* P_3^2 + 30a_{111} P_3^4} \tag{14.b}$$

$$r_{42} = \frac{\partial\left(\frac{1}{n_4^2}\right)}{\partial E_2} = \varepsilon_0\Big(\frac{8a_{123} P_3}{4a_{12}^* + 4a_{123} P_3^2} + \frac{4a_{13}^* P_3 + 8a_{112} P_3^3}{2a_1^* + 2a_{13}^* P_3^2 + 2a_{112} P_3^4} \tag{14.c}$$

$$+ \frac{4a_{13}^* + 24a_{112} P_3^2}{4a_{13}^* P_3 + 8a_{112} P_3^3}\Big)$$

We use the same procedure to derive the equations 9.a-c except the fact that the $P_1$ and $P_2$ are non-zero term in the orthorhombic phase.

## Acknowledgements

This work was partially supported by the Army Research Office under award no. W911NF-21-1-0327. The work at USC was also supported in part by ARO under award No. W911NF-19-1-0137, the National Science Foundation (NSF) through award DMR-2122071, and the Air Force Office of Scientific Research under contract FA9550-16-1-0335. Y. L. and J. R. acknowledge the Center for Advanced Research Computing (CARC) at the University of Southern California for providing computing resources that have contributed to the research results reported within this publication. URL: https://carc.usc.edu. The work at Washington University was also supported in part by NSF through awards DMR-1806147, DMR-2122070 and DMR-1931610. This work used the computational resources of the Extreme Science and Engineering Discovery Environment (XSEDE), which is supported by NSF grant ACI-1548562. The authors gratefully acknowledge Mr. Wente Li and Prof. Alex Demkov of UT Austin for feedback on the manuscript.

## Data Availability

The data that support the findings of this study are available from the corresponding authors upon reasonable request.

Page 23 of 31

# References

[1] P. K. Panda, *Review: Environmental Friendly Lead-Free Piezoelectric Materials*, J. Mater. Sci. **44**, 5049 (2009).

[2] Z. Sun, Z. Wang, Y. Tian, G. Wang, W. Wang, M. Yang, X. Wang, F. Zhang, and Y. Pu, *Progress, Outlook, and Challenges in Lead-Free Energy-Storage Ferroelectrics*, Adv. Electron. Mater. **6**, 1 (2020).

[3] J. Hao, W. Li, J. Zhai, and H. Chen, *Progress in High-Strain Perovskite Piezoelectric Ceramics*, Mater. Sci. Eng. R Reports **135**, 1 (2019).

[4] A. S. Bhalla, R. Guo, and R. Roy, *The Perovskite Structure - A Review of Its Role in Ceramic Science and Technology*, Mater. Res. Innov. **4**, 3 (2000).

[5] R. Weis and T. Gaylord, *Lithium Niobate: Summary of Physical Properties and Crystal Structure R.*, Appl. Phys. A Mater. Sci. Process. **37**, 191 (1985).

[6] D. G. Schlom, L. Q. Chen, X. Pan, A. Schmehl, and M. A. Zurbuchen, *A Thin Film Approach to Engineering Functionality into Oxides*, J. Am. Ceram. Soc. **91**, 2429 (2008).

[7] D. H. Kim and H. S. Kwok, *Pulsed Laser Deposition of BaTiO3 Thin Films and Their Optical Properties*, Appl. Phys. Lett. **67**, 1803 (1995).

[8] D. M. Gill, C. W. Conrad, G. Ford, B. W. Wessels, and S. T. Ho, *Thin-Film Channel Waveguide Electro-Optic Modulator in Epitaxial BaTiO3*, Appl. Phys. Lett. **71**, 1783 (1997).

[9] G. Yi, Z. Wu, and M. Sayer, *Preparation of Pb(Zr,Ti)O3 Thin Films by Sol Gel Processing: Electrical, Optical, and Electro-Optic Properties*, J. Appl. Phys. **64**, 2717 (1988).

[10] D. H. Reitze, E. Haton, R. Ramesh, S. Etemad, D. E. Leaird, T. Sands, Z. Karim, and A. R. Tanguay, *Electro-Optic Properties of Single Crystalline Ferroelectric Thin Films*, Appl. Phys. Lett. **63**, 596 (1993).

[11] R. A. McKee, F. J. Walker, and M. F. Chisholm, *Crystalline Oxides on Silicon: The First Five Monolayers*, Phys. Rev. Lett. **81**, 3014 (1998).

[12] S. R. Bakaul, C. R. Serrao, O. Lee, Z. Lu, A. Yadav, C. Carraro, R. Maboudian, R. Ramesh, and S. Salahuddin, *High Speed Epitaxial Perovskite Memory on Flexible Substrates*, Adv. Mater. **29**, 1605699 (2017).

[13] S. H. Baek and C. B. Eom, *Epitaxial Integration of Perovskite-Based Multifunctional Oxides on Silicon*, Acta Mater. **61**, 2734 (2013).

[14] S. Abel, F. Eltes, J. E. Ortmann, A. Messner, P. Castera, T. Wagner, D. Urbonas, A. Rosa, A. M. Gutierrez, D. Tulli, P. Ma, B. Baeuerle, A. Josten, W. Heni, D. Caimi, L. Czornomaz, A. A. Demkov, J. Leuthold, P. Sanchis, and J. Fompeyrine, *Large Pockels Effect in Micro- and Nanostructured Barium Titanate Integrated on Silicon*, Nat. Mater. **18**, 42 (2019).

[15] K. J. Kormondy, Y. Popoff, M. Sousa, F. Eltes, D. Caimi, M. D. Rossell, M. Fiebig, P. Hoffmann, C. Marchiori, M. Reinke, M. Trassin, A. A. Demkov, J. Fompeyrine, and S. Abel, *Microstructure and Ferroelectricity of $BaTiO_3$ Thin Films on Si for Integrated Photonics*, Nanotechnology **28**, 075706 (2017).

[16] C. Wang, M. Zhang, X. Chen, M. Bertrand, A. Shams-Ansari, S. Chandrasekhar, P. Winzer, and M. Lončar, *Integrated Lithium Niobate Electro-Optic Modulators Operating at CMOS-Compatible Voltages*, Nature **562**, 101 (2018).

[17] S. Abel, T. Stöferle, C. Marchiori, C. Rossel, M. D. Rossell, R. Erni, D. Caimi, M. Sousa, A. Chelnokov, B. J. Offrein, J. Fompeyrine, S. Abel, T. Sto, D. Caimi, M. Sousa, A. Chelnokov, B. J. Offrein, and J. Fompeyrine, *A Strong Electro-Optically Active Lead-Free Ferroelectric Integrated on Silicon*, Nat. Commun. **4**, 1671 (2013).

[18] H. Akazawa and M. Shimada, *Electro-Optic Properties of c-Axis Oriented LiNbO3 Films Grown on Si(100) Substrate*, Mater. Sci. Eng. B **120**, 50 (2005).

[19] T. Paoletta and A. A. Demkov, *Pockels Effect in Low-Temperature Rhombohedral BaTiO3*, Phys. Rev. B **103**, 14303 (2021).

[20] A. K. Hamze, M. Reynaud, J. Geler-Kremer, and A. A. Demkov, *Design Rules for Strong Electro-Optic Materials*, Npj Comput. Mater. **6**, (2020).

[21] K. D. Fredrickson, V. V. Vogler-Neuling, K. J. Kormondy, D. Caimi, F. Eltes, M. Sousa, J. Fompeyrine, S. Abel, and A. A. Demkov, *Strain Enhancement of the Electro-Optical Response in $BaTiO_3$ Films Integrated on Si(001)*, Phys. Rev. B **98**, 1 (2018).

[22] A. K. Hamze and A. A. Demkov, *First-Principles Study of the Linear Electro-Optical Response in Strained $SrTiO_3$*, Phys. Rev. Mater. **2**, 1 (2018).

[23] C. Paillard, S. Prokhorenko, and L. Bellaiche, *Strain Engineering of Electro-Optic Constants in Ferroelectric Materials*, Npj Comput. Mater. **5**, 1 (2019).

[24] M. DiDomenico and S. H. Wemple, *Oxygen-Octahedra Ferroelectrics. I. Theory of Electro-Optical and Nonlinear Optical Effects*, J. Appl. Phys. **40**, 720 (1969).

[25] S. H. Wemple and M. DiDomenico, *Oxygen-Octahedra Ferroelectrics. II. Electro-Optical and Nonlinear-Optical Device Applications*, J. Appl. Phys. **40**, 735 (1969).

[26] M. Veithen and P. Ghosez, *First-Principles Study of the Dielectric and Dynamical Properties of Lithium Niobate*, Phys. Rev. B **65**, 214302 (2002).

[27] M. Veithen and P. Ghosez, *Temperature Dependence of the Electro-Optic Tensor and Refractive Indices of BaTiO3 from First Principles*, Phys. Rev. B **71**, 132101 (2005).

[28] M. Veithen, X. Gonze, and P. Ghosez, *First-Principles Study of the Electro-Optic Effect in Ferroelectric Oxides*, Phys. Rev. Lett. **93**, 187401 (2004).

[29] M. Veithen, X. Gonze, and P. Ghosez, *Nonlinear Optical Susceptibilities, Raman Efficiencies, and Electro-Optic Tensors from First-Principles Density Functional Perturbation Theory*, Phys. Rev. B **71**, 125107Í‹ (2005).

[30] J. H. Qiu, J. N. Ding, N. Y. Yuan, X. Q. Wang, and Y. Zhou, *Film Thickness Dependence of Electro-Optic Effect in Epitaxial BaTiO 3 Thin Films*, Solid State Commun. **151**, 1344 (2011).

[31] Y. Lu and R. J. Knize, *Enhanced Dielectric and Electro-Optic Effects in Relaxor Oxide Heterostructured Superlattices*, J. Phys. D. Appl. Phys. **37**, 2432 (2004).

[32] J. Hiltunen, D. Seneviratne, R. Sun, M. Stolfi, H. L. Tuller, J. Lappalainen, and V. Lantto, *BaTiO3–SrTiO3 Multilayer Thin Film Electro-Optic Waveguide Modulator*, Appl. Phys. Lett. **89**, 242904 (2006).

[33] P. Chandra and P. B. Littlewood, *A Landau Primer for Ferroelectrics*, in *Physics of Ferroelectrics* (Springer Berlin Heidelberg, Berlin, Heidelberg, 2007), pp. 69–116.

[34] K. G. Lim, K. H. Chew, L. H. Ong, and M. Iwata, *Recent Advances in Application of Landau-Ginzburg Theory for Ferroelectric Superlattices*, Solid State Phenom. **232**, 169 (2015).

[35] K. H. Chew, *Recent Applications of Landau-Ginzburg Theory to Ferroelectric Superlattices: A Review*, Solid State Phenom. **189**, 145 (2012).

[36] L.-Q. Chen, *Phase-Field Method of Phase Transitions/Domain Structures in Ferroelectric Thin Films: A Review*, J. Am. Ceram. Soc. **91**, 1835 (2008).

[37] A. I. Khan, K. Chatterjee, J. P. Duarte, Z. Lu, A. Sachid, S. Khandelwal, R. Ramesh, C. Hu, and S. Salahuddin, *Negative Capacitance in Short-Channel FinFETs Externally Connected to an Epitaxial Ferroelectric Capacitor*, IEEE Electron Device Lett. **37**, 111 (2016).

[38] P. Marton, I. Rychetsky, and J. Hlinka, *Domain Walls of Ferroelectric BaTiO3 within the Ginzburg-Landau-Devonshire Phenomenological Model*, Phys. Rev. B - Condens. Matter Mater. Phys. **81**, (2010).

[39] G. Kresse and J. Furthmüller, *Efficient Iterative Schemes for Ab Initio Total-Energy Calculations Using a Plane-Wave Basis Set*, Phys. Rev. B **54**, 11169 (1996).

[40] D. Joubert, *From Ultrasoft Pseudopotentials to the Projector Augmented-Wave Method*, Phys. Rev. B - Condens. Matter Mater. Phys. **59**, 1758 (1999).

[41] R. O. Jones and O. Gunnarsson, *The Density Functional Formalism, Its Applications and Prospects*, Rev. Mod. Phys. **61**, 689 (1989).

[42] J. P. Perdew, K. Burke, and M. Ernzerhof, *Generalized Gradient Approximation Made Simple*, Phys. Rev. Lett. **77**, 3865 (1996).

[43] D. I. Bilc, R. Orlando, R. Shaltaf, G. M. Rignanese, J. Íñiguez, and P. Ghosez, *Hybrid Exchange-Correlation Functional for Accurate Prediction of the Electronic and Structural Properties of Ferroelectric Oxides*, Phys. Rev. B - Condens. Matter Mater. Phys. **77**, 1 (2008).

[44] Y. Zhang, J. Sun, J. P. Perdew, and X. Wu, *Comparative First-Principles Studies of Prototypical Ferroelectric Materials by LDA, GGA, and SCAN Meta-GGA*, Phys. Rev. B **96**, 35143 (2017).

[45] R. D. King-Smith and D. Vanderbilt, *Theory of Polarization of Crystalline Solids*, Phys. Rev. B **47**, 1651 (1993).

[46] A. K. M. I. Aroyo, J. M. Perez-Mato, D. Orobengoa, E. Tasci, G. de la Flor, *No Title*, "Crystallography Online Bilbao Crystallogr. Server" Bulg. Chem. Commun. **43**, 183 (2011).

[47] B. J. C. H. T. Stokes, D. M. Hatch, *ISOTROPY Software Suite*, Iso.byu.edu.

[48] X. Gonze, *A Brief Introduction to the ABINIT Software Package*, Zeitschrift Für Krist. - Cryst. Mater. **220**, 558 (2005).

[49] X. Gonze, B. Amadon, P.-M. Anglade, J.-M. Beuken, F. Bottin, P. Boulanger, F. Bruneval, D. Caliste, R. Caracas, M. Côté, T. Deutsch, L. Genovese, P. Ghosez, M. Giantomassi, S. Goedecker, D. R. Hamann, P. Hermet, F. Jollet, G. Jomard, S. Leroux, M. Mancini, S. Mazevet, M. J. T. Oliveira, G. Onida, Y. Pouillon, T. Rangel, G.-M. Rignanese, D. Sangalli, R. Shaltaf, M. Torrent, M. J. Verstraete, G. Zerah, and J. W. Zwanziger, *ABINIT: First-Principles Approach to Material and Nanosystem Properties*, Comput. Phys. Commun. **180**, 2582 (2009).

[50] N. Troullier and J. L. Martins, *Efficient Pseudopotentials for Plane-Wave Calculations. II. Operators for Fast Iterative Diagonalization*, Phys. Rev. B **43**, 8861 (1991).

[51] M. Zgonik, P. Bernasconi, M. Duelli, R. Schlesser, P. Günter, M. H. Garrett, D. Rytz, Y. Zhu, and X. Wu, *Dielectric, Elastic, Piezoelectric, Electro-Optic, and Elasto-Optic Tensors of BaTiO3 Crystals*, Phys. Rev. B **50**, 5941 (1994).

[52] F. Jona and G. Shirane, *Ferroelectric Crystals* (Wiley Online Library, 1962).

[53] Karin M. Rabe ; Charleshn ; Jean-Marc Triscone, *Physics of Ferroelectrics: Topics in Applied Physics* (2007).

[54] G. A. Samara, *Pressure and Temperature Dependence of the Dielectric Properties and Phase Transitions of the Ferroelectric Perovskites: Pbtio3 and Batio3*, Ferroelectrics **2**, 277 (1971).

[55] R. C. Miller and A. Savage, *TEMPERATURE DEPENDENCE OF THE OPTICAL PROPERTIES OF FERROELECTRIC LiNbO 3 AND LiTaO 3*, Appl. Phys. Lett. **9**, 169 (1966).

[56] J. L. Serving and F. Gervais, *Displacive-Type Phase Transition in LiTaO 3 and LiNbO 3*, Ferroelectrics **25**, 609 (1980).

[57] Y. L. Wang, A. K. Tagantsev, D. Damjanovic, N. Setter, V. K. Yarmarkin, and A. I. Sokolov, *Anharmonicity of BaTiO3 Single Crystals*, Phys. Rev. B - Condens. Matter Mater. Phys. **73**, 1 (2006).

[58] R. E. Cohen, *Origin of Ferroelectricity in Perovskite Oxides*, Nature **358**, 136 (1992).

[59] R. Resta and D. Vanderbilt, *Theory of Polarization: A Modern Approach*, in *Physics of Ferroelectrics: Topics in Applied Physics* (Springer Berlin Heidelberg, Berlin, Heidelberg, 2007), pp. 31–68.

[60] X. Lu, H. Li, and W. Cao, *Landau Expansion Parameters for BaTiO3*, J. Appl. Phys. **114**, (2013).

[61] M. J. Weber, *Handbook of Optical Materials*, Vol. 41 (CRC press, 2002).

[62] N. A. Pertsev, A. G. Zembilgotov, and A. K. Tagantsev, *Effect of Mechanical Boundary Conditions on Phase Diagrams of Epitaxial Ferroelectric Thin Films*, Phys. Rev. Lett. **80**, 1988 (1998).

[63] V. B. Shirokov, Y. I. Yuzyuk, B. Dkhil, and V. V. Lemanov, *Phenomenological Theory of Phase Transitions in Epitaxial BaTi O3 Thin Films*, Phys. Rev. B - Condens. Matter Mater. Phys. **75**, 1 (2007).

[64] R. W. Boyd, *Nonlinear Optics, Third Edition*, 3rd ed. (Academic Press, Inc., USA, 2008).

[65] J. A. Gonzalo and J. M. Rivera, *Ferroelectric Free Energy Expansion Coefficients from Double Hysteresis Loops*, Ferroelectrics **2**, 31 (1971).

[66] W. J. Merz, *Double Hysteresis Loop of BaTiO3 at the Curie Point*, Phys. Rev. **91**, 513 (1953).

[67] L.-Q. Chen, *Appendix A - Landau Free-Energy Coefficient*, in *Physics of Ferroelectrics: Topics in Applied Physics* (2007), pp. 363–370.

[68] J. Sun, R. C. Remsing, Y. Zhang, Z. Sun, A. Ruzsinszky, H. Peng, Z. Yang, A. Paul, U. Waghmare, X. Wu, M. L. Klein, and J. P. Perdew, *Accurate First-Principles Structures and Energies of Diversely Bonded Systems from an Efficient Density Functional*, Nat. Chem. **8**, 831 (2016).

[69] C. Herzog, G. Poberaj, and P. Günter, *Electro-Optic Behavior of Lithium Niobate at Cryogenic Temperatures*, Opt. Commun. **281**, 793 (2008).

[70] P. Bernasconi, M. Zgonik, and P. Gunter, *Temperature Dependence and Dispersion of Electro-Optic and Elasto-Optic Effect in Perovskite Crystals*, J. Appl. Phys. **78**, 2651 (1995).

[71] W. J. Merz, *Domain Formation and Domain Wall Motions in Ferroelectric BaTiO3 Single Crystals*, Phys. Rev. **95**, 690 (1954).

[72] E. A. Little, *Dynamic Behavior of Domain Walls in Barium Titanate*, Phys. Rev. **98**, 978 (1955).

[73] S. V Kalinin, S. Jesse, B. J. Rodriguez, Y. H. Chu, R. Ramesh, E. A. Eliseev, and A. N. Morozovska, *Probing the Role of Single Defects on the Thermodynamics of Electric-Field Induced Phase Transitions*, Phys. Rev. Lett. **100**, 155703 (2008).

[74] S. Jesse, B. J. Rodriguez, S. Choudhury, A. P. Baddorf, I. Vrejoiu, D. Hesse, M. Alexe, E. A. Eliseev, A. N. Morozovska, J. Zhang, L.-Q. Chen, and S. V Kalinin, *Direct Imaging of the Spatial and Energy Distribution of Nucleation Centres in Ferroelectric Materials*, Nat. Mater. **7**, 209 (2008).

[75] L. D. Landau and E. M. Lifshitz, *Electrodynamics of Continuous Media*, Second Edi (1984).

[76] J. Li, B. Nagaraj, H. Liang, W. Cao, C. H. Lee, R. Ramesh, J. Li, B. Nagaraj, H. Liang, W. Cao, C. H. Lee, and R. Ramesh, *Ultrafast Polarization Switching in Thin-Film Ferroelectrics Ultrafast Polarization Switching in Thin-Film Ferroelectrics*, **1174**, 3 (2007).

[77] A. K. Tagantsev, I. Stolichnov, N. Setter, J. S. Cross, and M. Tsukada, *Non-Kolmogorov-Avrami Switching Kinetics in Ferroelectric Thin Films*, 1 (2002).

[78] J. F. Scott, *A Review of Ferroelectric Switching*, Ferroelectrics **503**, 117 (2016).

[79] Y. W. Li, J. F. Scott, D. N. Fang, F. X. Li, Y. W. Li, J. F. Scott, D. N. Fang, and F. X. Li, *90-Degree Polarization Switching in BaTiO3 Crystals without Domain Wall Motion 90-Degree Polarization Switching in BaTiO 3 Crystals without Domain Wall Motion*, **232901**, 0 (2014).

[80] K. Laabidi, M. D. Fontana, M. Maglione, B. Jannot, and K. A. Müller, *Indications of Two Separate Relaxators in the Subphonon Region of Tetragonal BaTiO 3*, Europhys. Lett. **26**, 309 (1994).

[81] Y. K. Vayunandana Reddy and D. Mergel, *Frequency and Temperature-Dependent Dielectric Properties of BaTiO3 Thin Film Capacitors Studied by Complex Impedance Spectroscopy*, Phys. B Condens. Matter **391**, 212 (2007).

[82] M. Zhang and C. Deng, *Orientation and Electrode Configuration Dependence on Ferroelectric, Dielectric Properties of BaTiO3 Thin Films*, Ceram. Int. **45**, 22716 (2019).

[83] A. Piorra, A. Petraru, H. Kohlstedt, M. Wuttig, and E. Quandt, *Piezoelectric Properties of 0.5(Ba0.7Ca0.3TiO3) - 0.5[Ba(Zr0.2Ti0.8)O3] Ferroelectric Lead-Free Laser Deposited Thin Films*, J. Appl. Phys. **109**, 3 (2011).

[84] D. Xue, Y. Zhou, H. Bao, C. Zhou, J. Gao, and X. Ren, *Elastic, Piezoelectric, and Dielectric Properties of Ba(Zr$_{0.2}$Ti$_{0.8}$)O$_3$-50(Ba$_{0.7}$Ca$_{0.3}$)TiO$_3$ Pb-Free Ceramic at the Morphotropic Phase Boundary*, J. Appl. Phys. **109**, 054110 (2011).

[85] A. D. Dupuy, Y. Kodera, and J. E. Garay, *Unprecedented Electro-Optic Performance in Lead-Free Transparent Ceramics*, Adv. Mater. **28**, 7970 (2016).

[86] G. Keiser, *Optical Fiber Communications*, in *Wiley Encyclopedia of Telecommunications* (John Wiley & Sons, Inc., Hoboken, NJ, USA, 2003).